\documentclass[journal=jctcce,manuscript=article]{achemso}

\usepackage[version=3]{mhchem} % Formula subscripts using \ce{}
\usepackage{balance} 
\usepackage{rotating}
\usepackage{url}
\usepackage{tabularx}
\usepackage{multirow}
\usepackage{amsmath,amssymb}
\usepackage{mathabx}
\usepackage{pifont}
\usepackage{color, colortbl}
\usepackage{bm}% bold math
\usepackage[table]{xcolor}
\usepackage{dcolumn}
\usepackage{longtable}
\usepackage{nicefrac}
\usepackage{tablefootnote}
\usepackage{threeparttable}
\usepackage{tablefootnote}
\usepackage{makecell}
\AtBeginDocument{}
\usepackage{chemmacros}
\newcommand{\Lower}[1]{\smash{\lower 1.5ex \hbox{#1}}}

\newcommand{\mc}{\multicolumn{1}{c}}

\usepackage{soul}

\newcolumntype{d}[1]{D{.}{.}{#1}}

\author{Katharina Boguslawski}
\email{k.boguslawski@fizyka.umk.pl}
\affiliation[Institute of Physics, Nicolaus Copernicus University in Torun]
{Institute of Physics, Faculty of Physics, Astronomy and Informatics, Nicolaus Copernicus University in Torun, Grudziadzka 5, 87-100 Torun, Poland}
\alsoaffiliation[Faculty of Chemistry, Nicolaus Copernicus University in Torun]
{Faculty of Chemistry, Nicolaus Copernicus University in Torun, Gagarina 7, 87-100 Torun, Poland}
\phone{(+48)56 611 3338}

\title[]
 {Targeting doubly-excited states with Equation of Motion Coupled Cluster theory restricted to double excitations}
 
\abbreviations{AP1roG, LCCSD, EOM, pCCD}
\begin{document}
%%%%%%%%%%%%%%%%%%%%%%%%%%%%%%%%%%%%%%%%%%%%%%%%%%%%%%%%%%%%%%%%%%%%%
%% The manuscript does not need to include \maketitle, which is
%% executed automatically.  The document should begin with an
%% abstract, if appropriate.  If one is given and should not be, the
%% contents will be gobbled.
%%%%%%%%%%%%%%%%%%%%%%%%%%%%%%%%%%%%%%%%%%%%%%%%%%%%%%%%%%%%%%%%%%%%%
 
\begin{abstract}
The accurate description of doubly-excited states using conventional electronic structure methods is remarkably challenging, primarily because such excited states require the inclusion of doubly or higher excited configurations or the application of multi-reference methods.
We present a new approach to target electronically excited states that feature a double-electron transfer.
Our method uses the equation of motion (EOM) formalism with a pair coupled cluster doubles (pCCD) reference function, where dynamical correlation is accounted for by a linearized coupled cluster correction with singles and doubles (LCCSD).
Specifically, our proposed EOM-pCCD-LCCSD model represents a simplification of the conventional EOM-CCSD approach, where the electron-pair amplitudes of CCSD are tailored by pCCD.
The performance of EOM-pCCD-LCCSD is assessed for the lowest-lying excited states in \ce{CH^+} and all-trans polyenes.
In contrast to conventional EOM-CC methods with at most double excitations, EOM-pCCD-LCCSD predicts the right order of states in polyenes with excitation energies closest to experiment, outperforming even highly-accurate methods such as the density matrix renormalization group algorithm.
\end{abstract}
%%%%%%%%%%%%%%%%%%%%%%%%%%%%%%%%%%%%%%%%%%%%%%%%%%%%%%%%%%%%%%%%%%%%%%%%%%%%%%%%%%%%%%%%%%%%%%%%%%%%%%%%%%%%%%%%%%%%%%%%%%

%%%%%%%%%%%%%%%%%%%%%%%%%%%%%%%%%%%%%%%%%%%%%%%%%%%%%%%%%%%%%%%%%%%%%%%%%%%%%%%%%%%%%%%%%%%%%%%%%%%%%%%%%%%%%%%%%%%%%%%%%%
%%%%%%%%%%%%%%%%%%%%%%%%%%%%%%%%%%%%%%%%%%%%%%%%%%%%%%%%%%%%%%%%%%%%%%%%%%%%%%%%%%%%%%%%%%%%%%%%%%%%%%%%%%%%%%%%%%%%%%%%%%
\section{Introduction}
%%%%%%%%%%%%%%%%%%%%%%%%%%%%%%%%%%%%%%%%%%%%%%%%%%%%%%%%%%%%%%%%%%%%%%%%%%%%%%%%%%%%%%%%%%%%%%%%%%%%%%%%%%%%%%%%%%%%%%%%%%
%%%%%%%%%%%%%%%%%%%%%%%%%%%%%%%%%%%%%%%%%%%%%%%%%%%%%%%%%%%%%%%%%%%%%%%%%%%%%%%%%%%%%%%%%%%%%%%%%%%%%%%%%%%%%%%%%%%%%%%%%%
The efficient and reliable description of electronically excited states of atoms and molecules is gaining in importance in many areas of chemistry, physics, biology, and materials science. 
This trend promotes the development of new quantum chemistry methods dedicated to specifically model electronically excited states properties in large molecules and complex systems~\cite{gomes_rev_2012,doi:10.1002/wcms.99,kowalski-eom-review-2011,krylov-review,lischka2018multireference,kowalski-eom-2002,jorgensen-ch+}.
Such methods are, however, usually designed to accurately treat electronic excitations with a dominant transfer of one single electron. 
Excitation energies featuring a double-electron transfer are more problematic as their description requires the inclusion of higher order excitations or even a multi-reference treatment~\cite{kowalski-importance-of-triples}.
Both methodologies typically suffer from an unfavorable computational scaling, which limits their application to small model systems. 
To advance a reliable, but inexpensive description of doubly-excited states in large molecules, we have developed a new approach based on the computationally tractable pair coupled cluster doubles (pCCD) model~\cite{Limacher_2013,tamar-pcc}.
Specifically, the pCCD approach represents a simplified version of CCD, where only electron-pair excitations are kept in the cluster operator,
%The cluster operator $\hat{T}_{\rm p}$ takes, thus, the form
%%%%%%%%%%%%%%%%%%%%%%%%%%%%%%%%%%%%%%%%%%%%%%%%%%%%%%%%%%%%%%%%%%%%%%%%%%%%%%%%%%%%%%%%%%%%%%%%%%%%%%%%%%%%%%%%%%%%%%%%%%
%%%%%%%%%%%%%%%%%%%%%%%%%%%%%%%%%%%%%%%%%%%%%%%%%%%%%%%%%%%%%%%%%%%%%%%%%%%%%%%%%%%%%%%%%%%%%%%%%%%%%%%%%%%%%%%%%%%%%%%%%%
\begin{equation}\label{eq:ap1rog}
|{\rm pCCD}\rangle = \exp \left (  \sum_{i=1}^{\rm occ} \sum_{a=1}^{\rm virt} t_i^a a_a^{\dagger}  a_{\bar{a}}^{\dagger}a_{\bar{i}} a_{i}  \right )| \Phi_0 \rangle = e^{\hat{T}_{\rm p}} | \Phi_0 \rangle,
\end{equation}
%%%%%%%%%%%%%%%%%%%%%%%%%%%%%%%%%%%%%%%%%%%%%%%%%%%%%%%%%%%%%%%%%%%%%%%%%%%%%%%%%%%%%%%%%%%%%%%%%%%%%%%%%%%%%%%%%%%%%%%%%%
%%%%%%%%%%%%%%%%%%%%%%%%%%%%%%%%%%%%%%%%%%%%%%%%%%%%%%%%%%%%%%%%%%%%%%%%%%%%%%%%%%%%%%%%%%%%%%%%%%%%%%%%%%%%%%%%%%%%%%%%%%
where $a^\dagger_p$ and $a_p$ ($a^\dagger_{\bar{p}}$ and $a_{\bar{p}}$) are the electron creation and annihilation operators for $\alpha$ ($\beta$) electrons and $| \Phi_0 \rangle$ is some reference determinant. %(for instance, the Hartree--Fock (HF) determinant).
By construction, pCCD can only describe correlations restricted to electron pairs. The missing correlation effects that can be attributed to broken-pair states have to be included \textit{a posteriori}.
This can be achieved, for instance, using perturbation theory~\cite{Piotrus_PT2,AP1roG-PTX} or CC corrections~\cite{frozen-pccd,kasia-lcc}.
In the latter approach, the electronic wave function is written using an exponential ansatz with some cluster operator $\hat{T}= \sum_\nu t_\nu \tau_\nu$ ($\hat{\tau}$ is an excitation operator) and the pCCD wave function as reference state,
%%%%%%%%%%%%%%%%%%%%%%%%%%%%%%%%%%%%%%%%%%%%%%%%%%%%%%%%%%%%%%%%%%%%%%%%%%%%%%%%%%%%%%%%%%%%%%%%%%%%%%%%%%%%%%%%%%%%%%%%%%
        \begin{equation}\label{eq:cc}
            |\Psi \rangle = \exp({\hat{T}})  \vert {\rm pCCD} \rangle.
        \end{equation}
%%%%%%%%%%%%%%%%%%%%%%%%%%%%%%%%%%%%%%%%%%%%%%%%%%%%%%%%%%%%%%%%%%%%%%%%%%%%%%%%%%%%%%%%%%%%%%%%%%%%%%%%%%%%%%%%%%%%%%%%%%
$\hat{T}$ may contain single excitations $\hat{T}_1$, (non-pair) double excitations $\hat{T}_2^\prime$, and higher excitations~\cite{frozen-pccd}.
Note that the $\prime$ indicates the exclusion of pair excitations, that is, $\hat{T}_2 = \hat{T}_2^\prime + \hat{T}_{\rm p}$.
To facilitate the distinction between electron-pair (p) excitations and non-pair (np) excitations, we will label the excitation operators accordingly.
The CC correction can be further simplified by considering a linearized CC (LCC) ansatz.~\cite{kasia-lcc}
The cluster amplitudes $t_\nu$ are then determined by solving a linear set of coupled equations 
%%%%%%%%%%%%%%%%%%%%%%%%%%%%%%%%%%%%%%%%%%%%%%%%%%%%%%%%%%%%%%%%%%%%%%%%%%%%%%%%%%%%%%%%%%%%%%%%%%%%%%%%%%%%%%%%%%%%%%%%%%
        \begin{equation}\label{eq:lcc}
            \langle \Phi_\nu \vert (\hat{H}_N + [\hat{H}_N,\hat{T}_{\rm np}]) \vert {\rm pCCD} \rangle =  0,
        \end{equation}
%%%%%%%%%%%%%%%%%%%%%%%%%%%%%%%%%%%%%%%%%%%%%%%%%%%%%%%%%%%%%%%%%%%%%%%%%%%%%%%%%%%%%%%%%%%%%%%%%%%%%%%%%%%%%%%%%%%%%%%%%%
where the Baker--Campbell--Hausdorff expansion $e^{-\hat{T}_{\rm np}}\hat{H}_Ne^{\hat{T}_{\rm np}}  = \hat{H}_N + [\hat{H}_N,\hat{T}_{\rm np}] + \frac{1}{2} [ [\hat{H}_N,\hat{T}_{\rm np}], \hat{T}_{\rm np}] + \ldots$ has been truncated after the second term, $\hat{H}_N = \hat{H} - \langle \Phi_0 \vert \hat{H} \vert \Phi_0 \rangle$ is the quantum chemical Hamiltonian in its normal product form, and $| \Phi_\nu \rangle = \hat{\tau}_\nu | \Phi_0 \rangle$.
%If we substitute the exponential form of the pCCD wave function, eq.~\eqref{eq:lcc} can be brought into the familiar form of single-reference coupled cluster theory,
Using eq.~\eqref{eq:ap1rog}, eq.~\eqref{eq:lcc} can be brought into the familiar form of single-reference CC theory,
%%%%%%%%%%%%%%%%%%%%%%%%%%%%%%%%%%%%%%%%%%%%%%%%%%%%%%%%%%%%%%%%%%%%%%%%%%%%%%%%%%%%%%%%%%%%%%%%%%%%%%%%%%%%%%%%%%%%%%%%%%
        \begin{equation}\label{eq:pccdlcc}
            \langle \Phi_\nu \vert (\hat{H}_N + [\hat{H}_N,\hat{T}_{\rm np}] + [\hat{H}_N,\hat{T}_{\rm p}] + [ [\hat{H}_N,\hat{T}_{\rm np}],\hat{T}_{\rm p}]) \vert \Phi_0 \rangle =  0.
        \end{equation}
%%%%%%%%%%%%%%%%%%%%%%%%%%%%%%%%%%%%%%%%%%%%%%%%%%%%%%%%%%%%%%%%%%%%%%%%%%%%%%%%%%%%%%%%%%%%%%%%%%%%%%%%%%%%%%%%%%%%%%%%%%
%We should emphasize again that, by construction, $\hat{T}_{\rm np}$ contains excitations beyond electron pairs, \textit{i.e.}, $\hat{T}_{\rm np}=\hat{T}_1+\hat{T}_2^\prime$.
Thus, the hybrid pCCD-LCCSD (linearized coupled cluster singles and doubles) approach represents a simplification of CCSD and frozen-pair (fp)CCSD~\cite{frozen-pccd}, where all non-pair amplitudes enter linearly into the CC amplitudes equation eq.~\eqref{eq:pccdlcc}.
%In contrast to pCCD-LCCSD, conventional LCC theory with a HF reference function excludes the last term of eq.~\eqref{eq:pccdlcc}. %, which contains the direct coupling between pair and non-pair amplitudes.
%Thus, 
The $\hat{T}_{\rm p}$ operator introduces non-linear terms into the single and double amplitudes equations, which explicitly couple the pair amplitudes with all non-pair amplitudes.
%The energy contribution from the LCCSD correction can be calculated by projecting against the reference determinant of $\vert {\rm pCCD} \rangle$,
%%%%%%%%%%%%%%%%%%%%%%%%%%%%%%%%%%%%%%%%%%%%%%%%%%%%%%%%%%%%%%%%%%%%%%%%%%%%%%%%%%%%%%%%%%%%%%%%%%%%%%%%%%%%%%%%%%%%%%%%%%
%        \begin{equation}
%          \langle \Phi_0 \vert  (\hat{H}_N + [\hat{H}_N,\hat{T}_{\rm np}] ) \vert \Phi_0 \rangle = E^{\rm LCCSD},
%        \end{equation}
%%%%%%%%%%%%%%%%%%%%%%%%%%%%%%%%%%%%%%%%%%%%%%%%%%%%%%%%%%%%%%%%%%%%%%%%%%%%%%%%%%%%%%%%%%%%%%%%%%%%%%%%%%%%%%%%%%%%%%%%%%
%where higher-order terms do not contribute.
Most importantly, all proposed CC corrections on top of pCCD represent some tailored CC flavor, where the electron-pair amplitudes of the CC correction are taken from a previously optimized pCCD wave function.

Numerical studies suggest~\cite{kasia-lcc,AP1roG-PTX} that pCCD-LCCSD is a reliable wave function ansatz to model both static and dynamic electron correlation
%Furthermore, pCCD-LCCSD
and allows us to reach chemical accuracy ($\sim1$ kcal/mol) for many challenging systems.~\cite{AP1roG-PTX}
The pCCD-LCCSD method is, however, applicable to ground-state electronic structures only and must be extended to target excited states.
Recently, we presented the first extension of pCCD to describe excited-state electronic structures, where single excitations are treated approximately in the excited state model.~\cite{eompccd,eompccderratum}
The proposed EOM-pCCD+S ansatz can reliably describe singly-excited states when dynamic electron correlation effects are similar in both the ground and excited states.
If, however, correlation effects differ in the ground and excited states, the corresponding excitation energies will be inaccurate.
To improve the EOM-pCCD+S model, we will present an excited state model based on the pCCD-LCCSD reference function.
The proposed excited state ansatz will be assessed against the lowest-lying excited states of the \ce{CH^+} molecule and all-trans polyenes with up to 7 double bonds.

%%%%%%%%%%%%%%%%%%%%%%%%%%%%%%%%%%%%%%%%%%%%%%%%%%%%%%%%%%%%%%%%%%%%%%%%%%%%%%%%%%%%%%%%%%%%%%%%%%%%%%%%%%%%%%%%%%%%%%%%%%
%%%%%%%%%%%%%%%%%%%%%%%%%%%%%%%%%%%%%%%%%%%%%%%%%%%%%%%%%%%%%%%%%%%%%%%%%%%%%%%%%%%%%%%%%%%%%%%%%%%%%%%%%%%%%%%%%%%%%%%%%%
\section{Extending pCCD-LCCSD to model excited states}\label{sec:theory}
%%%%%%%%%%%%%%%%%%%%%%%%%%%%%%%%%%%%%%%%%%%%%%%%%%%%%%%%%%%%%%%%%%%%%%%%%%%%%%%%%%%%%%%%%%%%%%%%%%%%%%%%%%%%%%%%%%%%%%%%%%
%%%%%%%%%%%%%%%%%%%%%%%%%%%%%%%%%%%%%%%%%%%%%%%%%%%%%%%%%%%%%%%%%%%%%%%%%%%%%%%%%%%%%%%%%%%%%%%%%%%%%%%%%%%%%%%%%%%%%%%%%%
We will target excited states using the equation-of-motion (EOM) formalism, where excited states are modelled using a linear CI-type ansatz,
\begin{equation}
        \hat{R} = \sum_\mu c_\mu \hat{\tau}_\mu.
\end{equation}
The above sum runs over all excitations present in the cluster operator as well as the identity operator $\hat{\tau}_0$.
The $\hat{R}$ operator, then, generates the excited state by acting on the CC reference state,
\begin{equation}
        \vert \Psi \rangle = \hat{R} \exp(\hat{T}) | \Phi_0 \rangle = \sum_\mu {c_\mu \hat{\tau}_\mu} \exp(\hat{T}) | \Phi_0 \rangle.
\end{equation}
%Throughout this work, we will disregard any excitation properties, like dipole moments, and focus on excitation energies.
Since we will focus on excitation energies, we have to solve for the $\hat{R}$ amplitudes only.
Introducing the similarity transformed Hamiltonian in normal-product form $\hat{\mathcal{H}}_N = \exp{(-\hat{T})} \hat{H}_N \exp{(\hat{T})}$ and subtracting the equation for the CC ground state, we obtain the EOM-CC equations, % for the $\hat{R}$ amplitudes,
\begin{equation}\label{eq:eomcc}
        [\hat{\mathcal{H}}_N,\hat{R}] \vert \Phi_0 \rangle  = \omega \hat{R} \vert \Phi_0 \rangle,
\end{equation}
where $\omega=(E-E_0)$ are the excitation energies with respect to the CC ground state, $ \exp(\hat{T}) | \Phi_0 \rangle$.
As the pCCD-LCCSD cluster operator contains all single and double excitations, the $\hat{R}$ operator becomes $\hat{R}=\hat{R}_0+\hat{R}_1+\hat{R}_2$.
%we have to include all these excitations (as well as the identity operator $\hat{\tau}_0$) in the $\hat{R}$ operator, \textit{i.e.}, $\hat{R}=\hat{R}_0+\hat{R}_1+\hat{R}_2$.
Furthermore, the similarity transformed Hamiltonian of pCCD-LCCSD has a special form~\cite{kasia-lcc}.
Specifically, for all non-pair excitations, we have the linearized approximation
$\mathcal{\hat{H}}^{\rm np}_N = e^{-\hat{T}_{\rm p}-\hat{T}_{\rm np}} \hat{H}_N e^{\hat{T}_{\rm p}+\hat{T}_{\rm np}}
                            \approx \hat{H}_N + [\hat{H}_N,\hat{T}] + [[\hat{H},\hat{T}_{\rm np}],\hat{T}_{\rm p}] $,
while the electron-pair amplitudes are optimized in pCCD with
$\mathcal{\hat{H}}^{\rm p}_N = e^{-\hat{T}_{\rm p}} \hat{H}_N e^{\hat{T}_{\rm p}}=
                             \hat{H}_N + [\hat{H}_N,\hat{T}_{\rm p}] + \frac{1}{2} [[\hat{H},\hat{T}_{\rm p}],\hat{T}_{\rm p}]$.
Although the similarity transformed Hamiltonian differs for electron-pair and non-pair excitations, we have considered the most general form of the $\hat{R}$ operator ($\hat{R}=\hat{R}_0+\hat{R}_1+\hat{R}_2$) when calculating the excitation energies from eq.~\eqref{eq:eomcc}.
%%%%%%%%%%%%%%%%%%%%%%%%%%%%%%%%%%%%%%%%%%%%%%%%%%%%%%%%%%%%%%%%%%%%%%%%%%%%%%%%%%%%%%%%%%%%%%%%%%%%%%%%%%%%%%%%%%%%%%%%%%
%%%%%%%%%%%%%%%%%%%%%%%%%%%%%%%%%%%%%%%%%%%%%%%%%%%%%%%%%%%%%%%%%%%%%%%%%%%%%%%%%%%%%%%%%%%%%%%%%%%%%%%%%%%%%%%%%%%%%%%%%%
%\begin{figure*}[p] %[htb]
%\centering
%\includegraphics[width=1.0\linewidth]{figure1}
%\caption{
%Diagrammatic representation of the left-hand-side of the EOM-pCCD-LCCSD equations eq.~\eqref{eq:eomcc}.
%(a) Diagrammatic representation of the left-hand-side of the single-excitation equations ($\hat{R}_1$).
%(b) Diagrammatic representation of the left-hand-side of the double-excitation equations ($\hat{R}_2$) partitioned into different contributions, which are also contained in EOM-pCCD+S (i), EOM-LCCSD (i and ii), and EOM-CCSD (i,ii, and iii).
%}
%\label{fig:diagrams}
%\end{figure*}
%%%%%%%%%%%%%%%%%%%%%%%%%%%%%%%%%%%%%%%%%%%%%%%%%%%%%%%%%%%%%%%%%%%%%%%%%%%%%%%%%%%%%%%%%%%%%%%%%%%%%%%%%%%%%%%%%%%%%%%%%%
%%%%%%%%%%%%%%%%%%%%%%%%%%%%%%%%%%%%%%%%%%%%%%%%%%%%%%%%%%%%%%%%%%%%%%%%%%%%%%%%%%%%%%%%%%%%%%%%%%%%%%%%%%%%%%%%%%%%%%%%%%
The diagrammatic representation of the left-hand-side of the EOM-pCCD-LCCSD equations eq.~\eqref{eq:eomcc} are summarized in Fig.~S1 of the Supporting Information.

{
Finally, we should note that the orbital-pairing scheme as well as orbital optimization are crucial in pCCD calculations.
Specifically, size consistency can be recovered if the orbitals are optimized within pCCD.
Numerical studies \mbox{suggest~\cite{AP1roG-PTX}} that dynamic energy corrections on top of pCCD yield similar equilibrium properties in canonical Hartree--Fock and pCCD optimized natural orbitals.
Hence, orbital optimization can be omitted if (relative) energies or properties of equilibrium structures or molecular structures close to the equilibrium are sought.
For stretched bonds or in the vicinity of dissociation, orbital optimization might become important to reach spectroscopic accuracy due to size-consistency errors.
Since, however, the molecular orbitals are typically optimized for the ground state, the conventional pCCD orbital optimization protocol results in molecular orbitals that are biased toward the ground state.
Furthermore, (variational) orbital optimization usually results in symmetry-broken (localized) \mbox{orbitals~\cite{oo-ap1rog,tamar-pcc,ps2-ap1rog,ap1rog-jctc}} that prevent us from identifying excited states according to their (point group) symmetry.
To provide information about spatial symmetry, excited state calculations can either be performed within the canonical Hartree--Fock orbital basis or using orbitals that have been optimized imposing point-group symmetry.
In the following, we will investigate both choices for excited state calculations with a pCCD-LCCSD reference function. 
}

%%%%%%%%%%%%%%%%%%%%%%%%%%%%%%%%%%%%%%%%%%%%%%%%%%%%%%%%%%%%%%%%%%%%%%%%%%%%%%%%%%%%%%%%%%%%%%%%%%%%%%%%%%%%%%%%%%%%%%%%%%
%%%%%%%%%%%%%%%%%%%%%%%%%%%%%%%%%%%%%%%%%%%%%%%%%%%%%%%%%%%%%%%%%%%%%%%%%%%%%%%%%%%%%%%%%%%%%%%%%%%%%%%%%%%%%%%%%%%%%%%%%%
%\section{Computational details}\label{sec:compdetails}
%%%%%%%%%%%%%%%%%%%%%%%%%%%%%%%%%%%%%%%%%%%%%%%%%%%%%%%%%%%%%%%%%%%%%%%%%%%%%%%%%%%%%%%%%%%%%%%%%%%%%%%%%%%%%%%%%%%%%%%%%%
%%%%%%%%%%%%%%%%%%%%%%%%%%%%%%%%%%%%%%%%%%%%%%%%%%%%%%%%%%%%%%%%%%%%%%%%%%%%%%%%%%%%%%%%%%%%%%%%%%%%%%%%%%%%%%%%%%%%%%%%%%

%%%%%%%%%%%%%%%%%%%%%%%%%%%%%%%%%%%%%%%%%%%%%%%%%%%%%%%%%%%%%%%%%%%%%%%%%%%%%%%%%%%%%%%%%%%%%%%%%%%%%%%%%%%%%%%%%%%%%%%%%%
%%%%%%%%%%%%%%%%%%%%%%%%%%%%%%%%%%%%%%%%%%%%%%%%%%%%%%%%%%%%%%%%%%%%%%%%%%%%%%%%%%%%%%%%%%%%%%%%%%%%%%%%%%%%%%%%%%%%%%%%%%
\section{Targeting doubly-excited states in \ce{CH+} and all-trans polyenes}\label{sec:results}
%%%%%%%%%%%%%%%%%%%%%%%%%%%%%%%%%%%%%%%%%%%%%%%%%%%%%%%%%%%%%%%%%%%%%%%%%%%%%%%%%%%%%%%%%%%%%%%%%%%%%%%%%%%%%%%%%%%%%%%%%%
%%%%%%%%%%%%%%%%%%%%%%%%%%%%%%%%%%%%%%%%%%%%%%%%%%%%%%%%%%%%%%%%%%%%%%%%%%%%%%%%%%%%%%%%%%%%%%%%%%%%%%%%%%%%%%%%%%%%%%%%%%
Our first test system comprises the excited-state potential energy surfaces (PESs) of a benchmark system for which full configuration interaction (FCI) results are available.
Specifically, we focus on the lowest-lying excited states of the \ce{CH+} molecule, whose valence excited states contain significant biexcited components when the \ce{C-H} bond is stretched.
Theoretical studies showed that the conventional EOM-CCSD approach yields large errors in excitation energies and breaks the degeneracy of states in the dissociation limit.~\cite{krylov2000,kowalski2001}
Thus, the \ce{CH+} molecule represents an ideal test system to assess the accuracy of the simplified EOM-pCCD-LCCSD formalism.

%%%%%%%%%%%%%%%%%%%%%%%%%%%%%%%%%%%%%%%%%%%%%%%%%%%%%%%%%%%%%%%%%%%%%%%%%%%%%%%%%%%%%%%%%%%%%%%%%%%%%%%%%%%%%%%%%%%%%%%%%%
%%%%%%%%%%%%%%%%%%%%%%%%%%%%%%%%%%%%%%%%%%%%%%%%%%%%%%%%%%%%%%%%%%%%%%%%%%%%%%%%%%%%%%%%%%%%%%%%%%%%%%%%%%%%%%%%%%%%%%%%%%
\begin{figure}[tb] %[htb]
\centering
\includegraphics[width=0.99\linewidth]{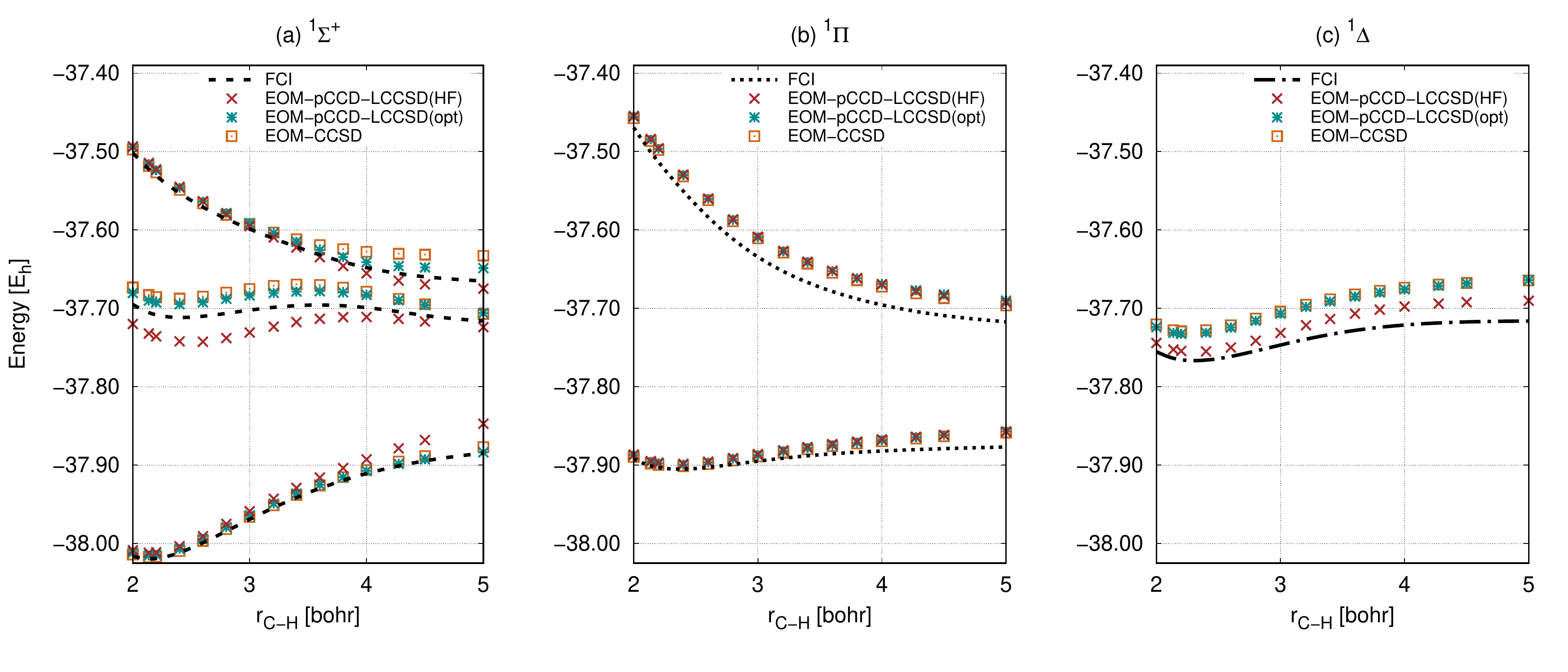}
\caption{\footnotesize
{
Potential energy curves for the lowest-lying states of the \mbox{\ce{CH+}} ion for FCI (black), pCCD-LCCSD (with (turquoise, \mbox{$\convolution$}) and without (red, \mbox{$\bm \times$}) orbital optimization), and EOM-CCSD (orange, \mbox{$\boxdot$}).
Dotted lines indicate FCI data, while points are used for EOM-CC data.
Only the three lowest \mbox{$^1\Sigma^+$} states (a), the two lowest \mbox{$^1\Pi$} (b), and the lowest \mbox{$^1\Delta$} state (c) are shown.
The total electronic energies and excitation energies can be found in Table S1 and S2 of the Supporting Information.
}
}
\label{fig:chplus}
\end{figure}
%%%%%%%%%%%%%%%%%%%%%%%%%%%%%%%%%%%%%%%%%%%%%%%%%%%%%%%%%%%%%%%%%%%%%%%%%%%%%%%%%%%%%%%%%%%%%%%%%%%%%%%%%%%%%%%%%%%%%%%%%%
%%%%%%%%%%%%%%%%%%%%%%%%%%%%%%%%%%%%%%%%%%%%%%%%%%%%%%%%%%%%%%%%%%%%%%%%%%%%%%%%%%%%%%%%%%%%%%%%%%%%%%%%%%%%%%%%%%%%%%%%%%

Fig.~\ref{fig:chplus} illustrates the PESs for the ground and lowest-lying excited states determined for FCI, EOM-CCSD, and EOM-pCCD-LCCSD and the three lowest-lying $^1\Sigma^+$ states (including the ground state), the two lowest-lying $^1\Pi$ states, and the lowest-lying $^1\Delta$ state.
Note that the first excited $^1\Sigma^+$ and $^1\Delta$ states feature a doubly-excited character, while the second $^1\Pi$ state has partial biexcited character.
Hence, we will focus our discussion on these particular states.
To facilitate our analysis, Table~\ref{tab:chplus} summarizes some selected error measures in excitation and total electronic energies with respect to FCI results.
These include the non-parallelity error (NPE $= \max\limits_{r_{\rm CH}}(|\Delta E_{r_{\rm CH}}|) - \min\limits_{r_{\rm CH}} (|\Delta E_{r_{\rm CH}}|) $ and $\Delta E_{r_{\rm CH}} = E^{\rm CC}_{r_{\rm CH}}-E^{\rm FCI}_{r_{\rm CH}}$), the maximum absolute error (MAE $=\max\limits_{r_{\rm CH}}(|\Delta E_{r_{\rm CH}}|)$), the mean error (ME $=\sum\limits_{r_{\rm CH}}\Delta E_{r_{\rm CH}}/N$), and the root mean error (RME $=\sqrt{\sum\limits_{r_{\rm CH}}\Delta E_{r_{\rm CH}}^2/N}$) determined for all C--H distances $r_{\rm CH}$ along each PES, where $N$ is the total number of points along the PES.
Note that for pCCD-LCCSD, we performed two different calculations: (i) using the canonical Hartree--Fock (HF) basis and (ii) using the optimized pCCD natural orbitals (opt) for the ground state molecule{ imposing \mbox{$C_{2h}$} symmetry (to label the excited states according to their symmetry)}.
While all $^1\Sigma^+$ states are better described within the optimized pCCD basis, the $^1\Pi$ and $^1\Delta$ states have smaller error measures using canonical HF orbitals.
Nonetheless, EOM-pCCD-LCCSD improves the description of all doubly-excited states and results in MAE of about 0.3 ($1 ^1\Delta$) to 0.4 ($2 ^1\Sigma$ and $2 ^1\Pi$) eV for excitation energies, while EOM-CCSD excitation energies yield an MAE of 1.2 ($1 ^1\Delta$) and about 0.6 ($2 ^1\Sigma$ and $2 ^1\Pi$) eV.
Furthermore, the corresponding RMEs are reduced from 1.1 eV to 0.15 eV ($1 ^1\Delta$) and from 0.5 to 0.3 eV ($2 ^1\Sigma$ and $2 ^1\Pi$).
Similar observation can be made for the total electronic energies.
Note, however, that all error measures for $E_{\rm el}$ increase by 0.1--0.4 eV.
In general, the excited-state PESs predicted by EOM-pCCD-LCCSD are qualitatively closer to the FCI reference curve and result in smaller NPEs (0.2 to 0.4 eV) compared to EOM-CCSD (0.3 to 0.78 eV).
Most importantly, EOM-pCCD-LCCSD(HF) can---at least partly---numerically restore the asymptotic degeneracy of the $1 ^1\Delta$ and $2 ^1\Pi$ state, fails, however, for the $2 ^1\Sigma$ state (those three states dissociate into C($^1$D) and H$^+$).
Specifically, the energy difference between the $1 ^1\Delta$ and $2 ^1\Pi$ state at 5.0 bohr is 0.066 eV for EOM-pCCD-LCCSD(HF) and increases to 0.865 eV for EOM-CCSD, while the $2 ^1\Sigma$ state lies 0.917 (1.163) eV below the $1 ^1\Delta$ state for EOM-pCCD-LCCSD (EOM-CCSD).
Orbital optimization within pCCD, however, breaks the asymptotic degeneracy of the $2 ^1\Sigma$, $1 ^1\Delta$, and $2 ^1\Pi$ states and results in similar errors as conventional EOM-CCSD.
Nonetheless, the proposed EOM-pCCD-LCCSD formalism significantly improves both excitation energies and PESs for excited states that feature a biexcited character compared to the conventional EOM-CCSD model.

\begin{table}[tb] %[htb]
\begin{center}
\caption{\footnotesize
Error measures [eV] for the ground and lowest-lying excited states of \ce{CH+} determined for pCCD-LCCSD (with and without orbital optimization) and EOM-CCSD with respect to FCI.~\cite{krylov2000}
The error measures are calculated for the total electronic energy $E_{\rm el}$ of each state and the corresponding excitation energies $\omega_{\rm el}$.
NPE: non-parallelity error;
%NPE(eq): NPE for $2 \leq r_{\rm CH} \leq 1.5r_e$ with $r_e= 2.13713$ bohr;
MAE: maximum absolute error;
ME: mean error;
RME: root mean error.
}\label{tab:chplus}
{\scriptsize
\begin{tabular}{ll|cccc|cccc|cccc} \hline\hline
& & \multicolumn{4}{c|}{EOM-pCCD-LCCSD(HF)} & \multicolumn{4}{c|}{EOM-pCCD-LCCSD(opt)} & \multicolumn{4}{c}{EOM-CCSD~\cite{kowalski2001}} \\  \hline
& State & NPE & MAE & ME & RME & NPE & MAE & ME & RME & NPE & MAE & ME & RME \\ \cline{2-14}
\multirow{6}{*}{$E_{\rm el}$}
 & X $^1\Sigma^+$  & 0.811 & 1.019 &  0.394 & 0.453 & 0.113 & 0.139 & 0.110 & 0.115 & 0.153 & 0.205 & 0.099 & 0.109 \\
 & $2 ^1\Sigma^+$  & 0.676 & 0.884 & -0.573 & 0.620 & 0.234 & 0.510 & 0.442 & 0.446 & 0.507 & 0.751 & 0.607 & 0.623 \\
 & $3 ^1\Sigma^+$  & 0.244 & 0.266 &  0.010 & 0.189 & 0.282 & 0.447 & 0.221 & 0.232 & 0.775 & 0.883 & 0.334 & 0.422 \\
 & $1 ^1\Pi$       & 0.372 & 0.530 &  0.298 & 0.321 & 0.396 & 0.540 & 0.275 & 0.301 & 0.420 & 0.498 & 0.226 & 0.261 \\
 & $2 ^1\Pi$       & 0.317 & 0.716 &  0.637 & 0.645 & 0.353 & 0.735 & 0.638 & 0.648 & 0.322 & 0.650 & 0.566 & 0.575 \\
 & $1 ^1\Delta$    & 0.406 & 0.707 &  0.481 & 0.506 & 0.572 & 1.426 & 1.114 & 1.126 & 0.459 & 1.408 & 1.177 & 1.186 \\
\hline \hline
\multirow{5}{*}{$\omega_{\rm el}$}
 & $2 ^1\Sigma^+$   & 0.411 & 1.229 & -0.967 & 0.974 & 0.124 & 0.373 & 0.332 & 0.334 & 0.628 & 0.668 & 0.508 & 0.539 \\
 & $3 ^1\Sigma^+$   & 1.282 & 1.282 & -0.383 & 0.555 & 0.382 & 0.421 & 0.110 & 0.150 & 0.622 & 0.678 & 0.235 & 0.317 \\
 & $1 ^1\Pi$        & 0.460 & 0.489 & -0.096 & 0.154 & 0.499 & 0.514 & 0.164 & 0.225 & 0.267 & 0.293 & 0.127 & 0.153 \\
 & $2 ^1\Pi$        & 0.400 & 0.431 &  0.243 & 0.317 & 0.456 & 0.710 & 0.528 & 0.544 & 0.288 & 0.563 & 0.467 & 0.476 \\
 & $1 ^1\Delta$     & 0.277 & 0.312 &  0.088 & 0.151 & 0.675 & 1.400 & 1.003 & 1.022 & 0.306 & 1.203 & 1.079 & 1.083 \\
\hline \hline
\end{tabular} 
}

\end{center}
\end{table}
%%%%%%%%%%%%%%%%%%%%%%%%%%%%%%%%%%%%%%%%%%%%%%%%%%%%%%%%%%%%%%%%%%%%%%%%%%%%%%%%%%%%%%%%%%%%%%%%%%%%%%%%%%%%%%%%%%%%%%%%%%
%%%%%%%%%%%%%%%%%%%%%%%%%%%%%%%%%%%%%%%%%%%%%%%%%%%%%%%%%%%%%%%%%%%%%%%%%%%%%%%%%%%%%%%%%%%%%%%%%%%%%%%%%%%%%%%%%%%%%%%%%%

Our second test system covers all-trans polyenes comprising two to seven double bonds.
All-trans polyenes are considerably challenging to model theoretically, primarily because doubly-excited configuration are required to accurately describe ground and excited states of longer polyenes~\cite{short-polyenes-review}.
Furthermore, the proper description of the two lowest-lying excited states poses a problem to both experiment and quantum chemistry approaches~\cite{Hudson1972,Schulten1976,Cave1987,Tavan1987,Cave1988,Buma1991,Orlandi1991,Serrano1993a,Serrano1993b,Krawczyk2000,Hsu2001,Wiberg2002,Starcke2006,polyeneLimit}.
%As the lowest-lying excited state has the same spatial symmetry as the ground state, the corresponding electronic transition is optically forbidden.
%Thus, the so-called first dark state $2^1A_g^-$ cannot be investigated using conventional electron spectroscopy.
%The first optically allowed state is the $1^1B_u^+$ state and lies energetically above the $2^1A_g^-$ state.~\cite{Hudson1972,Orlandi1991}
From a theoretical point of view, the lowest-lying excited states of all-trans polyenes have to be studied using highly-accurate electronic structure methods, like complete-active-space second-order perturbation theory (CASPT2) and the density-matrix renormalization-group algorithm (DMRG).
Such calculations predict the $2^1A_g^-$ state to be the lowest excited state and suggest a significant double excitation character with a dominant HOMO$^2\rightarrow$LUMO$^2$ transition.
%for carotenoids and polyene chromophores that play an important role in photoprocesses.
%The proper description of the two lowest-lying excited states poses a challenge to both experiment and quantum chemistry approaches~\cite{Hudson1972,Schulten1976,Cave1987,Tavan1987,Cave1988,Buma1991,Orlandi1991,Serrano1993a,Serrano1993b,Krawczyk2000,Hsu2001,Wiberg2002,Starcke2006,polyeneLimit}, especially because doubly-excited configuration are required to accurately model ground and excited states of longer polyenes.

%In this work, we will investigate the performance of various (approximate) EOM-CC models that are restricted to at most double excitations in predicting the lowest-lying excited states of all-trans polyenes of different chain lengths.
The excitation energies for the two lowest-lying vertical excited states of \ce{C4H6} to \ce{C14H16} obtained by various EOM-CC methods using the DFT-optimized structures of ref.~\citenum{eompccd} are summarized in Table~\ref{tab:dft}.
%The excitation energies obtained in the 6-31G basis are given in parenthesis next to the corresponding results calculated for the cc-pVDZ basis set.
%Note that the 6-31G basis set was used in CIS(D)~\cite{Starcke2006}, while a cc-pVDZ basis set was employed in MRMP calculations~\cite{Kurashige2004}.
%Since EOM-pCCD can only be used to target pair-excited states, only the excitation energies of the $2^1A_g^-$ state are shown in the Table.
Conventional multi-reference methods like CASSCF and MRMP predict the correct order of states. %, where the first dark states lies below the first bright state for longer polyene chain lengths.
As expected, EOM-CCSD overestimates the excitation energies of the $2^1A_g^-$ state and suggests that the dark state lies above the bright state by 0.9 eV.
Similar results are obtained from its linearized version EOM-LCCSD and the simple pCCD-based excited state model EOM-pCCD+S.
We should note that we encountered convergence difficulties in EOM-LCCSD for longer polyene chain lengths.
Thus, the corresponding excitation energies are not shown in the Table.
In contrast to the conventional EOM-CCSD method and its simplified variants, the EOM-pCCD-LCCSD approach predicts the first dark state to be the lowest-lying excited state.
Although being similar in computational cost as conventional EOM-CCSD, EOM-pCCD-LCCSD yields the correct order of states.
%The corresponding excitation energies lie between CASSCF and MRMP results.
We should emphasize that all EOM-CC methods perform equally good for the first bright state $1^1B_u^+$, whose state is mostly characterized by the single HOMO$^1\rightarrow$LUMO$^1$ transition.
Differences with respect to EOM-CCSD typically amount up to 0.3 eV, while the largest deviations are found for EOM-LCCSD (from 0.4 to 1.1 eV).
%The largest deviation from EOM-CCSD can be found for the linearized variant EOM-LCCSD, where differences in excitation energies increase for increasing polyene chain lengths (from 0.4 to 1.1 eV).
%For longer all-trans polyenes, we encountered convergence difficulties in EOM-LCCSD calculations
%Hence, the corresponding excitation energies are not given in Table~\ref{tab:dft}.

%%%%%%%%%%%%%%%%%%%%%%%%%%%%%%%%%%%%%%%%%%%%%%%%%%%%%%%%%%%%%%%%%%%%%%%%%%%%%%%%%%%%%%%%%%%%%%%%%%%%%%%%%%%%%%%%%%%%%%%%%%%%%%%%%%%%%%%%%%%%
%%%%%%%%%%%%%%%%%%%%% correlation energy plot  %%%%%%%%%%%%%%%%%%%%%%%%%%%%%%%%%%%%%%%%%%%%%%%%%%%%%%%%%%%%%%%%%%%%%%%%%%%%%%%%%%%%%%%%%%%%%
\begin{table}[tb] %[htb]
\centering
\caption{\footnotesize Vertical excitation energies [eV] of the two lowest-lying excited states in all-trans polyenes \ce{C4H6} to \ce{C14H16} calculated for various flavours of EOM-pCCD and different quantum chemistry methods.
Note that the 6-31G basis set was used in CIS(D), while the cc-pVDZ basis set was utilized in MRMP and all EOM-CC calculations. CASSCF was performed in a double-zeta basis set (see corresponding references).
%The excitation energies of EOM-pCCD and EOM-pCCD+S are determined for the cc-pVDZ basis set, while the corresponding results for the 6-31G basis set are given in parenthesis.
}\label{tab:dft}
\footnotesize
\begin{tabular}{l| ccccccc}
\hline
\hline
%& \multicolumn{5}{c|}{$2^1A_g^-$} & \multicolumn{4}{c}{$1^1B_u^+$} \\ \cline{2-10}
\ce{C=C}                  & 2             & 3              & 4             & 5             & 6             & 7             \\\hline
                          & \multicolumn{6}{c}{$2^1A_g^-$} \\\cline{2-7}
%EOM-pCCD~\cite{eompccd,eompccderratum}  &10.56\,(10.49) & 9.11\,(9.04)   & 8.11\,(8.02)  & 7.42\,(7.32)  & 6.93\,(6.82)  & 6.58\,(6.46)  \\
%EOM-pCCD+S~\cite{eompccd,eompccderratum}&7.45\,(7.37)   & 6.79\,(6.75)   & 6.15\,(6.09)  & 5.69\,(5.63)  & 5.37\,(5.29)  & 5.13\,(5.05)  \\
EOM-pCCD~\cite{eompccd,eompccderratum}    &10.56           & 9.11          & 8.11          & 7.42          & 6.93          & 6.58          \\
EOM-pCCD+S~\cite{eompccd,eompccderratum}  & 7.45           & 6.79          & 6.15          & 5.69          & 5.37          & 5.13          \\
EOM-pCCD-LCCSD            &6.26           & 5.29           & 4.58          & 4.11          & 3.79          & 3.57          \\
EOM-LCCSD                 &7.91           & 6.98           & 6.28          & 5.70          & 5.09          & \mc{--}       \\
EOM-CCSD                  &7.59           & 6.64           & 5.94          & 5.43          & 5.05          & 4.75          \\
CIS(D)~\cite{Starcke2006} &9.01           & 7.81           & 6.78          & 6.12          & 5.55          & 5.14          \\
MRMP~\cite{Kurashige2004} &6.31           & 5.10           & 4.26          & 3.68          & 3.19          & 2.80          \\
CASSCF~\cite{Nakayama}    &6.67           & 5.64           & 5.16          & 4.32          &\mc{--}        &\mc{--}        \\\hline
                          & \multicolumn{6}{c}{$1^1B_u^+$}  \\\cline{2-7}
%EOM-pCCD+S~\cite{eompccd,eompccderratum}&7.20\,(7.44)   & 5.98\,(6.16)   & 5.19\,(5.34)  & 4.62\,(4.75)  & 4.20\,(4.31)  & 3.87\,(3.97)  \\
EOM-pCCD+S~\cite{eompccd,eompccderratum}  & 7.20           & 5.98          & 5.19          & 4.62          & 4.20          & 3.87          \\
EOM-pCCD-LCCSD            &7.00           & 5.91           & 5.23          & 4.77          & 4.43          & 4.17          \\
EOM-LCCSD                 &7.24           & 6.27           & 5.70          & 5.37          & 5.28          &  --           \\
EOM-CCSD                  &6.86           & 5.74           & 5.03          & 4.53          & 4.16          & 3.87          \\
CIS(D)~\cite{Starcke2006} &8.09           & 6.78           & 5.95          & 5.43          & 5.00          & 4.70          \\
MRMP~\cite{Kurashige2004} &6.21           & 5.25           & 4.57          & 4.17          & 3.87          & 3.60          \\
CASSCF~\cite{Nakayama}    &7.73           & 7.06           & 6.62          & 6.37          &\mc{--}        &\mc{--}        \\
\hline
\hline
\end{tabular}
%\footnotetext[1]{Taken from Ref.~\cite{Starcke2006}}
%\footnotetext[2]{Taken from Ref.~\cite{Kurashige2004}}
%\footnotetext[3]{Taken from Ref.~\cite{Nakayama}}
\end{table} 
%%%%%%%%%%%%%%%%%%%%%%%%%%%%%%%%%%%%%%%%%%%%%%%%%%%%%%%%%%%%%%%%%%%%%%%%%%%%%%%%%%%%%%%%%%%%%%%%%%%%%%%%%%%%%%%%%%%%%%%%%%%%%%%%%%%%%%%%%%%%
%%%%%%%%%%%%%%%%%%%%%%%%%%%%%%%%%%%%%%%%%%%%%%%%%%%%%%%%%%%%%%%%%%%%%%%%%%%%%%%%%%%%%%%%%%%%%%%%%%%%%%%%%%%%%%%%%%%%%%%%%%%%%%%%%%%%%%%%%%%%

%%%%%%%%%%%%%%%%%%%%%%%%%%%%%%%%%%%%%%%%%%%%%%%%%%%%%%%%%%%%%%%%%%%%%%%%%%%%%%%%%%%%%%%%%%%%%%%%%%%%%%%%%%%%%%%%%%%%%%%%%%%%%%%%%%%%%%%%%%%%
%%%%%%%%%%%%%%%%%%%%% correlation energy plot  %%%%%%%%%%%%%%%%%%%%%%%%%%%%%%%%%%%%%%%%%%%%%%%%%%%%%%%%%%%%%%%%%%%%%%%%%%%%%%%%%%%%%%%%%%%%%
\begin{table}[tb] %[htb]
\centering
\caption{\footnotesize Vertical and adiabatic excitation energies [eV] of the two lowest-lying excited states in all-trans polyenes \ce{C10H12} to \ce{C14H16} calculated for various flavours of EOM-CC and DMRG.
%Different DMRG-optimized geometries have been used that have been optimized using different active spaces in DMRG calculations (see computational details and ref.~\citenum{dmrg-geom-opt}).
Note that all orbitals are active in all performed EOM-CC calculations.
%The DMRG reference data is taken from ref.~\citenum{dmrg-geom-opt}.
%Experimental data is taken from ref.~\citenum{Kohler}.
}\label{tab:dmrg}
\footnotesize
\begin{tabular}{ll| ccc | ccc}
\hline
\hline
& Excited state& \multicolumn{3}{c|}{$2^1A_g^-$} & \multicolumn{3}{c}{$1^1B_u^+$} \\ \cline{3-8}
                        & \ce{C=C}                  & 5    & 6    & 7    & 5    & 6    & 7 \\ \hline
\multirow{4}{*}{v}      &   EOM-pCCD+S~\cite{eompccd,eompccderratum}& 6.28 & 5.99 & 5.76 & 4.91 & 4.50 & 4.25 \\
                        &   EOM-pCCD-LCCSD          & 4.67 & 4.40 & 4.32 & 4.98 & 4.65 & 4.44 \\
                        &   EOM-LCCSD               & 6.11 & 5.68 & 5.39 & 5.42 & 5.15 & 4.98 \\
                        &   EOM-CCSD                & 5.79 & 5.43 & 5.22 & 4.78 & 4.41 & 4.19 \\
                        &   DMRG\cite{dmrg-geom-opt}& 5.43 & 4.76 & 4.64 & 5.35 & 4.98 & 4.66 \\ \hline
\multirow{4}{*}{a}      &   EOM-pCCD+S~\cite{eompccd,eompccderratum}& 5.18 & 4.61 & 4.44 & 5.20 & 4.30 & 4.04 \\
                        &   EOM-pCCD-LCCSD          & 3.47 & 2.77 & 1.85 & 5.08 & 4.38 & 4.15 \\
                        &   EOM-LCCSD               & 4.81 & 6.47 & 6.02 & 5.11 & 4.98 & 4.51 \\
                        &   EOM-CCSD                & 5.16 & 4.81 & 4.59 & 4.85 & 4.14 & 3.90 \\
                        &   DMRG\cite{dmrg-geom-opt}& 4.01 & 3.41 & 3.22 & 4.98 & 4.60 & 4.29 \\ \hline
%\multirow{4}{*}{v$_\pi$}&   EOM-pCCD+S~\cite{eompccd,eompccderratum}& 6.07 & 5.84 & 5.63 & 4.79 & 4.41 & 4.14 \\
%                        &   EOM-pCCD-LCCSD          & 4.45 & 4.23 & 4.12 & 4.90 & 4.59 & 4.37 \\
%                        &   EOM-LCCSD               & 5.95 & 5.54 & 5.15 & 5.39 & 5.15 & 5.01 \\
%                        &   EOM-CCSD                & 5.64 & 5.32 & 5.08 & 4.68 & 4.35 & 4.11 \\
%                        &   DMRG\cite{dmrg-geom-opt}& 4.51 & 4.15 & 3.91 & 5.77 & 5.41 & 5.16 \\ \hline
%\multirow{4}{*}{a$_\pi$}&   EOM-pCCD+S~\cite{eompccd,eompccderratum}& 4.85 & 4.60 & 4.45 & 4.63 & 4.26 & 3.99 \\
%                        &   EOM-pCCD-LCCSD 			& 2.97 & 2.75 & 1.96 & 4.70 & 4.40 & 4.18 \\
%                        &   EOM-LCCSD      			& --\footnotemark & 6.45 & 6.06 & 5.86 & 4.29 & 4.65 \\
%                        &   EOM-CCSD       			& 5.06 & 4.78 & 4.57 & 4.48 & 4.15 & 3.91 \\
%                        &   DMRG\cite{dmrg-geom-opt}& 3.36 & 2.99 & 2.73 & 5.49 & 5.13 & 4.87 \\ \hline
\multicolumn{2}{c|}{Exp.\cite{Kohler}}              & 3.03 & 2.69 & 2.44 & 3.57 & 3.31 & 3.12 \\
\hline
\hline
\end{tabular}
%\footnotetext[1]{Convergence difficulties}
\end{table} 
%%%%%%%%%%%%%%%%%%%%%%%%%%%%%%%%%%%%%%%%%%%%%%%%%%%%%%%%%%%%%%%%%%%%%%%%%%%%%%%%%%%%%%%%%%%%%%%%%%%%%%%%%%%%%%%%%%%%%%%%%%%%%%%%%%%%%%%%%%%%
%%%%%%%%%%%%%%%%%%%%%%%%%%%%%%%%%%%%%%%%%%%%%%%%%%%%%%%%%%%%%%%%%%%%%%%%%%%%%%%%%%%%%%%%%%%%%%%%%%%%%%%%%%%%%%%%%%%%%%%%%%%%%%%%%%%%%%%%%%%%
The influence of relaxation effects of the molecular structure on the excitation energies are summarized in Table~\ref{tab:dmrg}.
Note that the vertical and adiabatic excitation energies are calculated for the DMRG-optimized molecular geometries and hence do not equal the corresponding adiabatic excitation energies of each EOM-CC method.
We should note that two different active spaces have been used in DMRG calculations.
Since the EOM-CC excitation energies are similar for both DMRG structures, only one set of excitation energies is presented in the Table (see also Table S3 of the Supporting Information).
%The experimentally determined excitation energies for the two lowest excited states are given in the Table for comparison.
As observed for the DFT-optimized structures, all studied EOM-CC methods yield similar (vertical and adiabatic) excitation energies for the first bright state.
%The largest differences with respect to EOM-CCSD are again found for the EOM-LCCSD approach and amount up to 1 eV.
%Furthermore, all EOM-CC models predict similar (vertical and adiabatic) excitation energies for $1^1B_u^+$ for both investigated DMRG-optimized molecular structures, while the corresponding DMRG results differ by 0.4-0.5 eV.
In general, EOM-LCCSD and DMRG yield excitation energies that differ most from experimental reference data (1.5 to 2.0 eV), followed by EOM-pCCD-LCCSD (approximately 1.2 eV) and EOM-CCSD (approximately 1 eV).
%The best agreement is achieved in conventional EOM-CCSD calculations, where the theoretically predicted excitations energies deviate by approximately 1 eV from experimental data.
%The most unsteady performance is observed for EOM-LCCSD (differences with respect to EOM-CCSD are between 0.1 to 1 eV).
%While for some molecular structures and chain lengths, the excitation energy of $1^1B_u^+$ is similar to EOM-CCSD results, it can be overestimated by approximately 1 eV compared to EOM-CCSD results and by about 2.3 eV compared to experiment.
%We should note that EOM-pCCD+S slightly outperforms EOM-pCCD-LCCSD for the $1^1B_u^+$ state.
%However, the observed differences are typically smaller than 0.2 eV.

In contrast to the first bright state, the accurate description of the first dark state is more challenging.
While EOM-pCCD+S, EOM-LCCSD, and EOM-CCSD yield the wrong order of states, %where the $2^1A_g^-$ state lies above the $1^1B_u^+$ state,
DMRG and EOM-pCCD-LCCSD correctly predict the dark state to be the lowest-lying excited state.
%Specifically, %if dynamical electron correlation is accounted for on top of the pCCD wave function (as in pCCD-LCCSD), 
%the LCCSD correction lowers the excitation energies of $2^1A_g^-$ by about 1.5 eV (vertical excitations) and up to 2.5 eV (adiabatic excitations), respectively.
Compared to experimental results, EOM-CCSD overestimates excitation energies by 2.8 eV (vertical) and 2.1 eV (adiabatic), respectively, while DMRG yields differences between 2-2.4 eV (vertical) and 0.3-1.0 eV (adiabatic), respectively.
Most importantly, the EOM-pCCD-LCCSD excitation energies deviate less from experimental data (1.7-1.9 eV for vertical, 0.06-0.5 eV for adiabatic excitations). % : vertical excitation energies are overestimated by 1.7-1.9 eV, whereas adiabatic excitation energies differ by 0.06-0.5 eV from experimental results.
%Specifically, the adiabatic excitation energies of EOM-pCCD-LCCSD are closest to experimental values for \ce{C10H12} and \ce{C12H14} ($\Delta E(2^1A_g^-) = 0.06$ eV), while for \ce{C14H16} differences increase to 0.5-0.6 eV.
The large differences for \ce{C14H16} might originate from the fact that the corresponding ground and excited state molecular structures have not been optimized for the pCCD-LCCSD wave function or that pCCD-LCCSD is insufficient to describe the electron correlation effects in ground and excited states.
{
Furthermore, all ground and excited state calculations have been performed using the cc-pVDZ basis set to allow for a direct comparison to DMRG reference calculations.
For a better comparison to experiment, larger basis sets (of triple-zeta quality and augmented functions) should be applied.
Nonetheless, despite of the small size of the basis set and the errors in excitation energies, EOM-pCCD-LCCSD significantly outperforms conventional EOM-CC methods restricted to at most double excitations and predicts the right order of states in all-trans polyenes.}

%%%%%%%%%%%%%%%%%%%%%%%%%%%%%%%%%%%%%%%%%%%%%%%%%%%%%%%%%%%%%%%%%%%%%%%%%%%%%%%%%%%%%%%%%%%%%%%%%%%%%%%%%%%%%%%%%%%%%%%%%%
%%%%%%%%%%%%%%%%%%%%%%%%%%%%%%%%%%%%%%%%%%%%%%%%%%%%%%%%%%%%%%%%%%%%%%%%%%%%%%%%%%%%%%%%%%%%%%%%%%%%%%%%%%%%%%%%%%%%%%%%%%
\begin{figure}[t] %[htb]
\centering
\includegraphics[width=1.0\linewidth]{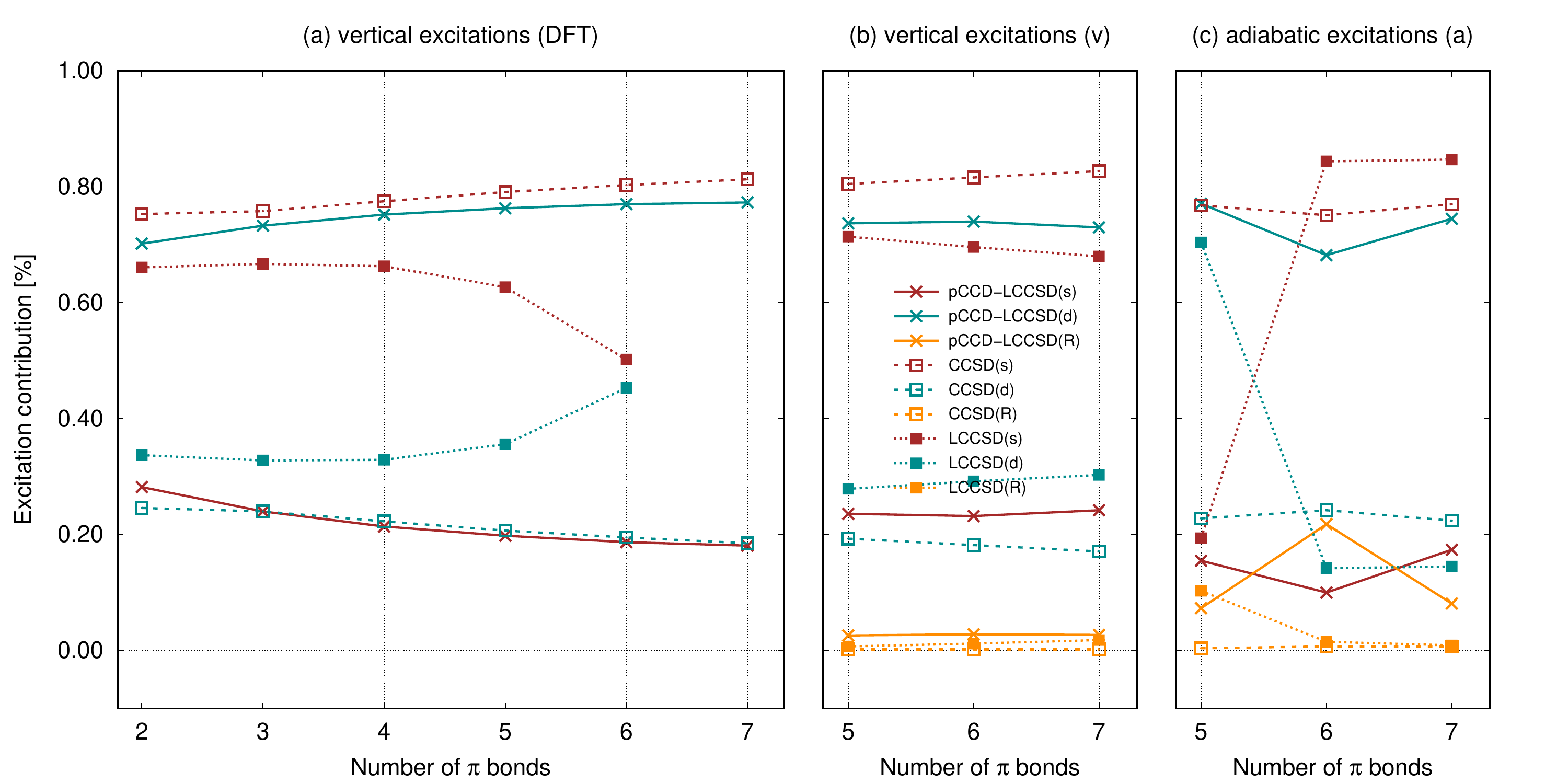}
\caption{\footnotesize
Contribution of doubly-excited (turquoise) and singly-excited (red) configurations as well as of the reference state (orange) for the $2^1A_g^-$ state calculated for various EOM-CC flavours using (a) the DFT-optimized structures, (b) and (c) the DMRG-optimized structures.
{$\bm \times$}: EOM-pCCD-LCCSD; 
{$\Box$}: EOM-CCSD;
\ding{110}: EOM-LCCSD.
}
\label{fig:character}
\end{figure}
%%%%%%%%%%%%%%%%%%%%%%%%%%%%%%%%%%%%%%%%%%%%%%%%%%%%%%%%%%%%%%%%%%%%%%%%%%%%%%%%%%%%%%%%%%%%%%%%%%%%%%%%%%%%%%%%%%%%%%%%%%
%%%%%%%%%%%%%%%%%%%%%%%%%%%%%%%%%%%%%%%%%%%%%%%%%%%%%%%%%%%%%%%%%%%%%%%%%%%%%%%%%%%%%%%%%%%%%%%%%%%%%%%%%%%%%%%%%%%%%%%%%%

The good performance of EOM-pCCD-LCCSD compared to the remaining EOM-CC models can be explained by analyzing the contributions of doubly- and singly-excited configurations, where % as shown in Fig.~\ref{fig:character}(a).
%It is well understood that the $2^1A_g^-$ state features three main configurations, which are the dominant doubly-excited $b_g\rightarrow a_u^*$ (HOMO$^2\rightarrow$LUMO$^2$), the singly-excited $b_g \rightarrow b_g^*$ (HOMO$\rightarrow$LUMO+1), and the singly-excited $a_u \rightarrow a_u^*$ (HOMO$-$1$\rightarrow$LUMO)~\cite{Schulten1976,Tavan1987}.
%Specifically, the doubly-excited $b_g\rightarrow a_u^*$ configuration comprises the largest weight.
%Furthermore,
the overall weight of all doubly-excited configurations may increase to approximately 80 \% for longer polyene chain lengths.~\cite{Starcke2006}
Fig.~\ref{fig:character}(a) shows the weight of singly-excited and doubly-excited configurations of the $2^1A_g^-$ state with respect to the number of double bonds (DFT-optimized structures) obtained by EOM-CCSD, its linearized version EOM-LCCSD, and the pair-amplitude tailored EOM-pCCD-LCCSD variant.
Both EOM-CCSD and EOM-LCCSD predict the first dark state to feature 70 to 80 \% singly-excited configurations. % with increasing weights for increasing polyene chain lengths.
%Specifically, for EOM-CCSD, the contribution of singly-excited configurations gradually rises for increasing polyene chain lengths, exceeding 80 \% for \ce{C14H16}, while the weight of doubly-excited configurations steadily decreases, falling below 20 \% for \ce{C14H16}.
For shorter polyene chain lengths, EOM-LCCSD behaves similar to EOM-CCSD. %, where the first dark state is dominated by singly-excited configurations.
If the number of double bonds exceeds 4, the contribution of singly-excited configurations rapidly decreases, while the weight of doubly-excited configurations significantly increases.
For \ce{C12H14}, both types of configurations have similar weights of approximately 50 \%.
%Note that for longer polyene chain lengths, we encountered convergence difficulties in EOM-LCCSD and hence the corresponding data is not shown in the Figure.
By contrast, EOM-pCCD-LCCSD correctly predicts the individual weights of singly- and doubly-excited configurations.
For an increasing number of $\pi$ bonds, the contribution of doubly-excited configurations gradually accumulates approaching 80 \%, while the weight of singly-excited configurations decreases accordingly.
%We should note that the relative weight of singly- and doubly-excited configurations is reverse for EOM-CCSD and EOM-pCCD-LCCSD.
To conclude, the character of the first dark state is appropriately described by the proposed EOM-pCCD-LCCSD model and lies energetically below the first bright state.
Similar trends can be observed for the DMRG-optimized structures  (see Fig.~\ref{fig:character}(b) and (c)). %and when the molecular geometry is allowed to relax.
%Fig.~\ref{fig:character}(b) and (c) shows the corresponding weights for the vertical (b) and adiabatic (c) $2^1A_g^-$ state for EOM-CCSD, EOM-LCCSD, and EOM-pCCD-LCCSD.
%Note that similar results have been obtained for both DMRG active spaces and hence only one set of results is shown in the Figure.
%While the weight of singly-excited configurations present in the vertically excited $2^1A_g^-$ state increases for EOM-CCSD for increasing polyene chain lengths, it gradually decreases for its linearized variant EOM-LCCSD.
%The reverse is true for doubly-excited configurations.
%For the adiabatically excited $2^1A_g^-$ state, the contribution of singly-excited configurations grows steadily with the number of double bonds for both EOM-CCSD and EOM-LCCSD.
%This large weight of singly-excited configurations in the adiabatic $2^1A_g^-$ state explains the poor performance of EOM-LCCSD for the adiabatic excitation energies of longer all-trans polyenes.
Specifically, EOM-LCCSD predicts larger weights of singly-excited configurations in the adiabatic $2^1A_g^-$ state, which may explain the poor performance of EOM-LCCSD for adiabatic excitation energies, while the contribution of singly-excited configurations in EOM-pCCD-LCCSD decreases from 20 \% to about 10 \% if the excited state structure is relaxed, while the weight of the reference state increases to 10-20 \%.
{
We should emphasize that EOM-LCCSD yields an adiabatically excited \mbox{$2^1A_g^-$ state of \ce{C10H12}} that features similar excitation contributions as obtained within EOM-pCCD-LCCSD, while the corresponding excitation energy lies between EOM-pCCD-LCCSD and EOM-CCSD results.
However, this rather good performance of EOM-LCCSD has been observed only for this particular molecular geometry and might be coincidental.
For all investigated all-trans polyenes with an increasing number of double bonds, LCCSD and EOM-LCCSD featured slow convergence in contrast to EOM-pCCD-LCCSD and EOM-CCSD calculations.
Furthermore, the excitation contributions of the EOM-LCCSD excited states are more sensitive to the molecular geometry than those obtained in EOM-pCCD-LCCSD and EOM-CCSD calculations.
}
%For EOM-pCCD-LCCSD, the contributions of doubly-excited configurations is similar for both the vertical and adiabatic state and accounts for approximately 80 \%.
%The weight of singly-excited configurations, however, decreases from 20 \% to about 10 \% if the excited state structure is relaxed, while the weight of the reference state significantly increases to 10-20 \%.
%Nonetheless, EOM-pCCD-LCCSD predicts the $2^1A_g^-$ state to contain approximately 80 \% of double excitation character for both vertical and adiabatic excitations.

%%%%%%%%%%%%%%%%%%%%%%%%%%%%%%%%%%%%%%%%%%%%%%%%%%%%%%%%%%%%%%%%%%%%%%%%%%%%%%%%%%%%%%%%%%%%%%%%%%%%%%%%%%%%%%%%%%%%%%%%%%
%%%%%%%%%%%%%%%%%%%%%%%%%%%%%%%%%%%%%%%%%%%%%%%%%%%%%%%%%%%%%%%%%%%%%%%%%%%%%%%%%%%%%%%%%%%%%%%%%%%%%%%%%%%%%%%%%%%%%%%%%%
\section{Conclusions}\label{sec:conclusion}
%%%%%%%%%%%%%%%%%%%%%%%%%%%%%%%%%%%%%%%%%%%%%%%%%%%%%%%%%%%%%%%%%%%%%%%%%%%%%%%%%%%%%%%%%%%%%%%%%%%%%%%%%%%%%%%%%%%%%%%%%%
%%%%%%%%%%%%%%%%%%%%%%%%%%%%%%%%%%%%%%%%%%%%%%%%%%%%%%%%%%%%%%%%%%%%%%%%%%%%%%%%%%%%%%%%%%%%%%%%%%%%%%%%%%%%%%%%%%%%%%%%%%
We have presented a new excited state model based on the pCCD wave function using the equation-of-motion formalism.
Our method represents a simplified version of EOM-CCSD where the electron-pair amplitudes are tailored by the pCCD cluster amplitudes and non-pair amplitudes enter only linearly into the CC and EOM equations.
%Specifically, the coupled cluster reference state is the pCCD-LCCSD reference function, where the electron-pair amplitudes are pre-optimized within pCCD and all remaining cluster amplitudes (singles and non-pair doubles) are optimized using a linearized coupled cluster ansatz.
%The special form of the pCCD-LCCSD ansatz entails that the amplitude equations contain both linear terms in all non-pair amplitudes and non-linear terms due to the coupling of the electron-pair and non-pair amplitudes.
%The corresponding EOM approach, thus, differs from the simple EOM-LCCSD model as it contains the coupling terms between pCCD and LCCSD.
The proposed EOM-pCCD-LCCSD model has been benchmarked against excitation energies that feature a doubly-excited character. %and excited-state PESs for the \ce{CH^+} molecule as well as excitation energies of all-trans polyenes containing 2 to 7 $\pi$ bonds.
In general, EOM-pCCD-LCCSD allows for an improved description of such excited states including at most double excitations in the CC ansatz.
Specifically, errors in excitation energies are reduced to 0.06--0.6 eV with respect to FCI or experimental reference data.
{
We should emphasize that pCCD calculations are commonly combined with an orbital optimization protocol (for the ground state) to recover size-consistency, while post-pCCD calculations are performed within the pCCD-optimized orbital basis.
The corresponding natural orbitals are typically symmetry-broken (localized), which prevents us from identifying excited states according to their spatial symmetry, and biased toward the ground state.
Symmetry-breaking can be prevented by imposing (point-group) symmetry in the orbital-optimization protocol.
}
To further improve both excitation energies and PESs{ and to eliminate the orbital bias toward the ground state}, the EOM-pCCD-LCCSD model can be combined with an orbital optimization protocol for excited states.
This is currently being investigated in our laboratory.

%%%%%%%%%%%%%%%%%%%%%%%%%%%%%%%%%%%%%%%%%%%%%%%%%%%%%%%%%%%%%%%%%%%%%%%%%%%%%%%%%%%%%%%%%%%%%%%%%%%%%%%%%%%%%%%%%%%%%%%%%%
%%%%%%%%%%%%%%%%%%%%%%%%%%%%%%%%%%%%%%%%%%%%%%%%%%%%%%%%%%%%%%%%%%%%%%%%%%%%%%%%%%%%%%%%%%%%%%%%%%%%%%%%%%%%%%%%%%%%%%%%%%
\section*{Computational Methodology}\label{sec:compdetails}
%%%%%%%%%%%%%%%%%%%%%%%%%%%%%%%%%%%%%%%%%%%%%%%%%%%%%%%%%%%%%%%%%%%%%%%%%%%%%%%%%%%%%%%%%%%%%%%%%%%%%%%%%%%%%%%%%%%%%%%%%%
%%%%%%%%%%%%%%%%%%%%%%%%%%%%%%%%%%%%%%%%%%%%%%%%%%%%%%%%%%%%%%%%%%%%%%%%%%%%%%%%%%%%%%%%%%%%%%%%%%%%%%%%%%%%%%%%%%%%%%%%%%
The molecular structures of all investigate all-trans polyenes containing 2 to 7 $\pi$-bonds were taken from ref.~\citenum{eompccd} (DFT-optimized structures, vertical excitations) and ref.~\citenum{dmrg-geom-opt} (DMRG-optimized structures, vertical and adiabatic excitations).
%Furthermore, the DMRG geometries were optimized using two different active spaces as discussed in ref.~\citenum{dmrg-geom-opt}: an active space containing only $\pi$-electrons (abbreviates as $v_\pi$ and $a_\pi$) and an active space comprising both $\sigma$- and $\pi$-electrons (denoted by $v$ and $a$, respectively).
%Similar to EOM-pCCD+S calculations, the molecular orbital basis was not optimized within the pCCD method (see also ref.~\citenum{eompccd} for more details).
For direct comparison with DMRG reference data, we used the cc-pVDZ basis set of Dunning~\cite{basis_dunning} in all calculations.
For the \ce{CH^+} ion, the basis set was taken from ref.~\citenum{olsen1989}.
All pCCD, EOM-pCCD-LCCSD, and EOM-LCCSD calculations were performed with our in-house quantum-chemistry code employing restricted canonical Hartree--Fock orbitals (\ce{CH^+} and polyenes) and pCCD natural orbitals optimized for the pCCD ground state reference function (\ce{CH^+}{, imposing $C_{2h}$ point group symmetry}) as molecular orbital basis.
%The lowest-lying excitation energies were determined employing a modified Davidson algorithm.
The EOM-CCSD calculations were performed with the \textsc{Molpro} program suite~\cite{molpro2012,molpro-wires}.

%%%%%%%%%%%%%%%%%%%%%%%%%%%%%%%%%%%%%%%%%%%%%%%%%%%%%%%%%%%%%%%%%%%%%%%%%%%%%%%%%%%%%%%%%%%%%%%%%%%%%%%%%%%%%%%%%%%%%%%%%%
%%%%%%%%%%%%%%%%%%%%%%%%%%%%%%%%%%%%%%%%%%%%%%%%%%%%%%%%%%%%%%%%%%%%%%%%%%%%%%%%%%%%%%%%%%%%%%%%%%%%%%%%%%%%%%%%%%%%%%%%%%
\section*{Acknowledgement}
%%%%%%%%%%%%%%%%%%%%%%%%%%%%%%%%%%%%%%%%%%%%%%%%%%%%%%%%%%%%%%%%%%%%%%%%%%%%%%%%%%%%%%%%%%%%%%%%%%%%%%%%%%%%%%%%%%%%%%%%%%
%%%%%%%%%%%%%%%%%%%%%%%%%%%%%%%%%%%%%%%%%%%%%%%%%%%%%%%%%%%%%%%%%%%%%%%%%%%%%%%%%%%%%%%%%%%%%%%%%%%%%%%%%%%%%%%%%%%%%%%%%%
K.B.~acknowledges financial support from a Marie-Sk\l{}odowska-Curie Individual Fellowship project no.~702635--PCCDX and a scholarship for outstanding young scientists from the Ministry of Science and Higher Education.
Calculations have been carried out using resources provided by Wroclaw Centre for Networking and Supercomputing (http://wcss.pl), grant no.~412.

%%%%%%%%%%%%%%%%%%%%%%%%%%%%%%%%%%%%%%%%%%%%%%%%%%%%%%%%%%%%%%%%%%%%%%%%%%%%%%%%%%%%%%%%%%%%%%%%%%%%%%%%%%%%%%%%%%%%%%%%%%
%%%%%%%%%%%%%%%%%%%%%%%%%%%%%%%%%%%%%%%%%%%%%%%%%%%%%%%%%%%%%%%%%%%%%%%%%%%%%%%%%%%%%%%%%%%%%%%%%%%%%%%%%%%%%%%%%%%%%%%%%%
\section*{Supporting Information}\label{sec:si}
%%%%%%%%%%%%%%%%%%%%%%%%%%%%%%%%%%%%%%%%%%%%%%%%%%%%%%%%%%%%%%%%%%%%%%%%%%%%%%%%%%%%%%%%%%%%%%%%%%%%%%%%%%%%%%%%%%%%%%%%%%
%%%%%%%%%%%%%%%%%%%%%%%%%%%%%%%%%%%%%%%%%%%%%%%%%%%%%%%%%%%%%%%%%%%%%%%%%%%%%%%%%%%%%%%%%%%%%%%%%%%%%%%%%%%%%%%%%%%%%%%%%%
Diagrammatic representation of EOM-pCCD-LCCSD, EOM-pCCD-LCCSD total electronic energies and excitation energies for the lowest-lying excited states of \ce{CH+}, and vertical and adiabatic excitation energies for \ce{C10H12} to \ce{C14H16} using all DMRG-optimized structures.

\bibliography{literature}
\end{document}